\documentclass[entropy,article,submit,moreauthors,pdftex,10pt,a4paper]{mdpi} 

\firstpage{1} 
\articlenumber{x}
\doinum{10.3390/------}
\pubvolume{xx}
\pubyear{2016}
\copyrightyear{2016}
\externaleditor{Academic Editor: name}
\history{Received: date; Accepted: date; Published: date}
\pdfoutput=1

 \theoremstyle{mdpi}
 \newcounter{thm}
 \newcounter{ex}
 \newcounter{re}
 \hypersetup{final}

 \theoremstyle{mdpidefinition}

\newcommand{\argmin}{\arg\!\min}

\usepackage{algorithm}
\usepackage{algorithmic}
\usepackage[T1]{fontenc}
\usepackage[utf8]{inputenc}
\usepackage{subcaption}

\floatname{algorithm}{Procedure}

\Title{Graph Energies of Egocentric Networks and Their Correlation with Vertex Centrality Measures}

\Author{Mikołaj Morzy$^{1,2}$* and Tomasz Kajdanowicz$^3$}

\address{%
$^1$ Institute of Computer Science, Poznań University of Technology, Piotrowo 2, 60-965 Poznań, Poland \\
$^2$ Center for Artificial Intelligence and Machine Learning CAMIL, Poznan University of Technology \\
$^3$ Department of Computational Intelligence, Wrocław University of Science and Technology, Wybrzeże Wyspiańskiego 27, 50-370 Wrocław, Poland\\
}
\corres{Correspondence: Mikolaj.Morzy@put.poznan.pl}

\abstract{Graph energy is the energy of the matrix representation of the graph, where the energy of a matrix is the sum of singular values of the matrix. Depending on the definition of a matrix, one can contemplate graph energy, Randi\'c energy, Laplacian energy, distance energy, and many others. Although theoretical properties of various graph energies have been investigated in the past in the areas of mathematics, chemistry, physics, or graph theory, these explorations have been limited to relatively small graphs representing chemical compounds or theoretical graph classes with strictly defined properties. In this paper we investigate the usefulness of the concept of graph energy in the context of large, complex networks. We show that when graph energies are applied to local egocentric networks, the values of these energies correlate strongly with vertex centrality measures. In particular, for some generative network models graph energies tend to correlate strongly with the betweenness and the eigencentrality of vertices. As the exact computation of these centrality measures is expensive and requires global processing of a network, our research opens the possibility of devising efficient algorithms for the estimation of these centrality measures based only on local information.}


\begin{document}

%%%%%%%%%%%%%%%%%%%%%%%%%%%%%%%%%%%%%%%%%%

\section{Introduction}
\label{sec:introduction}

Matrix energy is a concept well established in mathematics~\cite{biggs1993algebraic}, with several practical applications in other areas of science, such as physics and chemistry~\cite{cvetkovic1980spectra, kier2012molecular}. Since matrices are ubiquitous in the field of complex networks, it is natural to consider the usefulness of matrix energies in the context of networks. Complex networks, and social networks in particular, exhibit several interesting topological characteristics (dynamics, heterogeneity, high transitivity, relatively short average distances between vertices, degree associativity) and these characteristics vary significantly between individual vertices of a network. 

The term \emph{egocentric network} denotes a subnetwork consisting of a focal vertex, called \emph{ego}, and its direct neighbors, together with edges connecting these vertices. The main motivation behind our research is an attempt to use energy of egocentric network matrix as a meaningful feature for describing topological properties of the network. Our rationale is the following: connections within the egocentric network of the focal vertex $v$ define the the way in which any abstract resource (information, influence, importance) circulates in the vicinity of $v$. High singular values of the egocentric network matrix (and, in consequence, large energy of the egocentric network) characterize vertices which have high \emph{capacity}, i.e., vertices which can maintain a large amount of the resource in circulation, without having the resource leak out of the egocentric network. Recall that an \emph{eigenvector} of a linear transformation is a vector, whose direction does not change upon that linear transformation (in other words, the eigenvector of a linear transformation changes only by a scalar factor). If a linear transformation is represented by a square matrix $M$, then the column vector $\vec{v}$ is the eigenvector of $M$ if the following holds: $M \vec{v} = \lambda \vec{v}$, and $\lambda$ is a scalar value called the \emph{eigenvalue} of $M$. Thus, if one interprets the egocentric network adjacency matrix as a linear transformation describing the possible flow of a resource in the vicinity of the vertex $v$, large eigenvalues of that egocentric network adjacency matrix mean that the egocentric network of $v$ can circulate large amounts of a resource in a stable way (without changing the assignment of the resource to vertices, i.e., without changing the direction of the eigenvector). We have decided to explore the usefulness of the concept of egocentric network energy in characterizing individual vertices and see how this measure varies among vertices in networks representing different topologies and/or generative models.

To measure this variability, which can be interpreted in terms of the energy dispersion across the network, we employ the entropy. We also compute traditional centrality metrics for each vertex (degree, betweenness, closeness, eigencentrality) and compare these metrics with energies of the corresponding egocentric network.  We find surprising correlations between some of centrality metrics, which leads us to believe that matrix energies of egocentric networks can be very useful in describing topological positions of vertices.

In order to make our claims substantial, we perform experiments on four different generative network models. 

\begin{itemize}
    \item \textbf{Erd\"os-R\'enyi random network model}: a simple random network generator producing networks with binomial degree distribution,
    \item \textbf{Watts-Strogatz small network model}: a network generator producing networks with uniform degree distribution and very high local clustering coefficients,
    \item \textbf{Holme-Kim preferential attachment model}: a network generator producing networks with power law degree distributions and high local clustering coefficients,
    \item \textbf{Waxman geometric random network model}: an example of a geometric model producing random networks with grid-like structure.
\end{itemize}

We select generative network models in a way which covers a vast spectrum of possible network topologies, rather than choosing models that have strong practical applications. Although empirical networks that conform to the Erd\"os-R\'enyi random network model are rare, the model itself is an important abstraction and has nice analytical properties. Holme-Kim model is essentially identical to the far more popular Albert-Barab\'asi model~\cite{barabasi1999emergence}, but it also produces more realistic distributions of the local clustering coefficient. Finally, Waxman geometric model is a simple abstraction of a wide spectrum of networks with the underlying grid-like structures. Examples of such networks include transportation networks (e.g. railroad network), or infrastructure networks (e.g. power grid network, water supply network, Internet router network). Of course, each generative network model has several parameters which influence the characteristics of generated network instances, more detailed description of each generative network model is presented in Section~\ref{sec:network.models}. For each generative model we produce hundreds of instances of networks to better cover the space of possible network configurations.

Our main findings can be summarized as follows:

\begin{itemize}
    \item Graph energy and Laplacian energy correlate very strongly with several vertex features, in particular, these energies seem to agree with vertex degree, betweenness, and eigencentrality.
    \item Vertex closeness is notoriously difficult to estimate based on matrix energies.
    \item In all of the examined generative network models and for all considered matrix energies, there seems to be no consistent correlation of any graph energy with the local clustering coefficient.
    \item The entropy of all matrix energies is similar for a given network model, with Randi\'c energy being the most unstable across the spectrum of possible values of the model's parameter.
    \item High correlation of Graph energy and Laplacian energy with eigencentrality and betweenness suggests, that it is possible to devise methods for estimating the values of eigencentrality and betweenness based on local vertex energy instead of a costly global computation over the entire network.
\end{itemize}

\subsection{Related Work}

Graph energies~\cite{9781461442196} have been researched for many years in the domains of chemistry~\cite{cvetkovic1980spectra}, physics, mathematics~\cite{bernstein2005matrix}, and complex networks~\cite{van2010graph}. The branch of mathematics in which graph energies are developed is known as the \emph{algebraic graph theory}~\cite{cvetkovic2009introduction,godsil2001g}. Graph energies are matrix energies of various graph representations~\cite{gutman2001energy, Nikiforov2007a} and can be defined for any symmetric graph matrix~\cite{Consonni}. In the domain of mathematical chemistry they are known as \emph{spectral indices}. A spectral index is either a single eigenvalue, or a function of a set of eigenvalues of a matrix (this set of eignevalues is known as \emph{matrix spectrum}). For instance, one can use the sum of absolute eigenvalues, the sum of positive eigenvalues, the maximum eigenvalue, the minimum eigenvalue, the maximum absolute eigenvalue, the diameter of a spectrum (the difference between the maximum and the minimum eigenvalue), etc. Depending on the type of symmetric graph matrix, one can distinguish between \emph{graph energy}~\cite{nikiforov2016beyond} defined over the adjacency matrix, \emph{Randi\'c energy}~\cite{Randic1975} defined over the Randi\'c matrix, \emph{Laplacian energy}~\cite{Gutman2006, Merris1995} defined over the Laplacian matrix. Other types of graph matrices result in further spectral graph descriptors, such as \emph{Burden eigenvalues} defined for the modified connectivity matrix, \emph{incidence graph energy}~\cite{Gutman2009}, \emph{distance graph energy}~\cite{cash1995heats,Indulal2008}, or \emph{path energy}~\cite{Randic1975}. Each of these energies can be further quantified using a particular functional formula over the set of eigenvalues. Matrix energy is usually defined as the sum of absolute values of matrix eigenvalues, but other indices have been proposed, such as the Lovasz-Pelikan index~\cite{lovasz1973eigenvalues} (leading eigenvalue), VAA1 index~\cite{balaban1991topological} (sum of positive eigenvalues), quasi-Winer index~\cite{mohar1993novel} (sum of eigenvalue reciprocals for the Laplacian matrix), or the Estrada index~\cite{estrada2005subgraph,estrada2007statistical} (weighted number of the number of closed walks in the graph). Here we only reviewed a limited list of references but it must be mentioned that there are several excellent books on the subject of spectral graph theory, including \cite{Li2012, chung1997spectral}.

Matrix spectra and matrix energies find surprising applications in many areas of computer science~\cite{spielman2007spectral}. For instance, singular values of a matrix can be used to re-construct a matrix from a small set of incomplete entries~\cite{candes2010power,candes2009exact}, a problem common in recommender systems. Another application is graph drawing, where eigenvalues alone provide enough information to produce nice graph visualizations~\cite{hall1970r}. Graph eigenvalues have been successfully used to generate graph partitioning~\cite{fiedler1973algebraic,donath1972algorithms}. For many years graph eigenvalues were used to approximate the distribution of random walks over graphs~\cite{lovasz1993random}, with the PageRank algorithm being the most prominent example. Graph eigenvalues have found surprising applications in graph theory with respect to the graph coloring problem~\cite{wilf1967eigenvalues} and independent set problem~\cite{trevisan2012max}. Yet another area of application of matrix energies, far beyond the scope of this overview, is systems structural complexity. A comprehensive work on this subject can be found in~\cite{sinha2014structural}.

Graph and Randi\'c energies attract the most attention of the scientific community due to their direct applicability in molecular chemistry~\cite{li2008survey, Liu2008}. Several discoveries of properties of Randi\'c indexes and Randi\'c matrices have been made over the last years~\cite{Gutman2014, bozkurt2012randi, clark2000general}. The same can be said about the investigation of various properties of Laplacian energy in the context of random walks~\cite{zhou2008sum, stevanovic2009more, aleksic2008upper, li2009some, Gutman2006}. We feel that the concept of graph energy deserves much more attention from the social network analysis community and we intend to showcase the advantages of graph energies as vertex centrality measures. To the best of our knowledge there have been no previous works on the properties of graph energies computed over egocentric networks of vertices and their relationship to other vertex centrality measures.

The paper is organized as follows. Section~\ref{sec:basic.definitions} introduces basic definitions, matrix energies, and centrality measures. We present generative network models used in our experiments in Section~\ref{sec:network.models}. We describe our experiments and discuss the results in Section~\ref{sec:experiments}. The paper concludes in Section~\ref{sec:conclusions} with a brief summary.

\section{Basic Definitions}
\label{sec:basic.definitions}

\begin{figure}[h]
    \centering
    \includegraphics[width=0.5\textwidth]{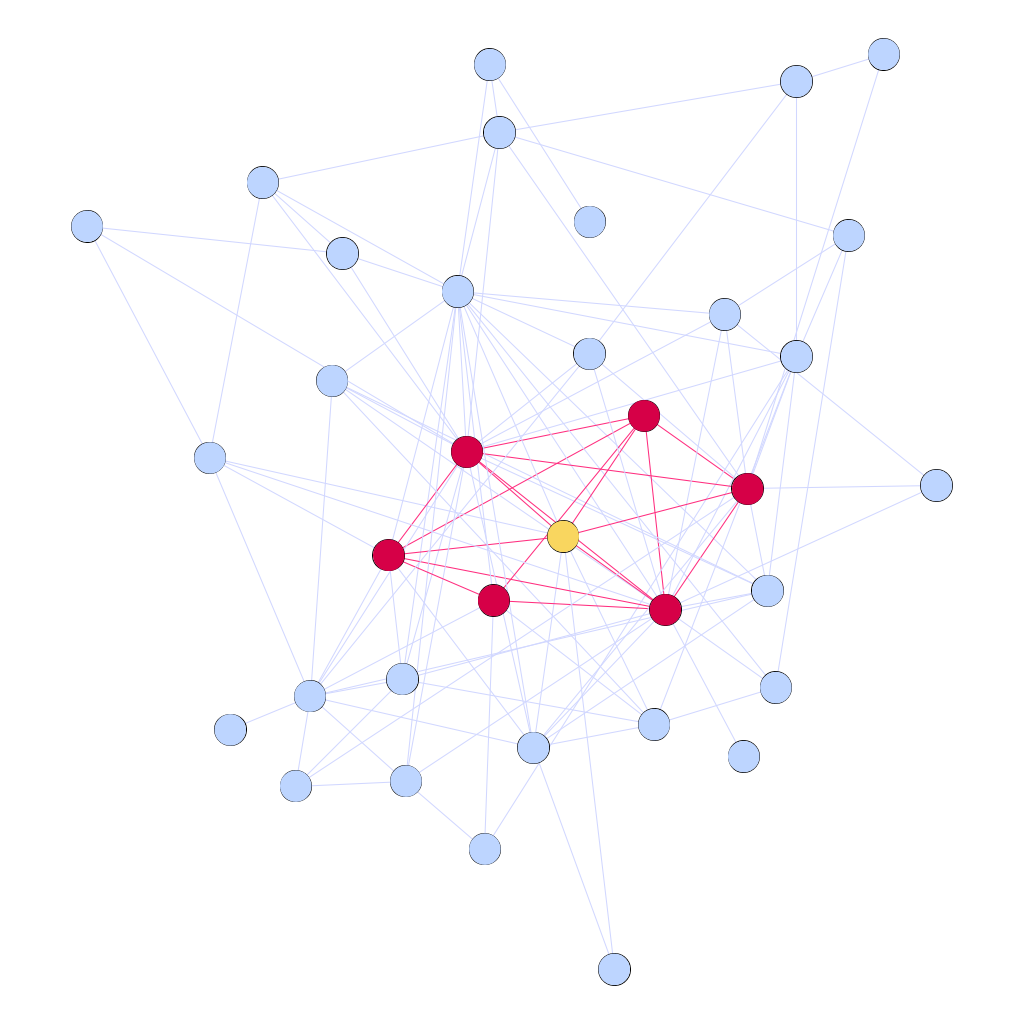}
    \caption{Egocentric network}
    \label{fig:ego.network}
\end{figure}

A complex network is a graph $G = \langle V, E \rangle$, where $V = \{ v_1, \ldots, v_n \}$ is the set of vertices, and $E = \{ \{v_i, v_j\} : v_i, v_j \in V \}$ is the set of edges, where each edge is an unordered pair of vertices from the set $V$ (in our experiments we consider undirected networks). The egocentric network of the vertex $v_i$, denoted $G_i=\langle V_i, E_i \rangle$, consists of the vertex $v_i$, all vertices adjacent to it, and all edges between those vertices, $V_i = \{v_i\} \cup \{v_j \in V : \{v_i, v_j\} \in E\}$, $E_i = \{ \{v_j,v_k\} \in E : v_j, v_k \in V_i \}$. Figure~\ref{fig:ego.network} presents an example of an egocentric network (red vertices and edges) of an ego (yellow vertex) within a larger network structure.

\subsection{Centrality Measures}

Centrality measures are functionals defined over the topology of a network which assign a numerical scalar to each vertex of a network. To capture different network and vertex characteristics, many such measures have been introduced~\cite{newman2018networks, wasserman1994social}. In this paper we utilize four of the most popular centrality measures (degree, betweenness, closeness, eigencentrality), but we note that many other such functionals exist, e.g., Katz centrality~\cite{katz1953new}, Bonacich power~\cite{bonacich1987power}, eccentricity~\cite{hage1995eccentricity}, etc.

\subsubsection{Degree}
The degree of the vertex $v_i$ is the number of edges adjacent to $v_i$:

\[ C_D(v_i) = \left| \{ v_j \in V : \{v_i, v_j\} \in E \} \right| \]

\subsubsection{Betweenness}
Let $p(v_i,v_j) = \langle v_i, v_{i+1}, \ldots, v_{j-1}, v_j \rangle$ be a sequence of vertices in which any two consecutive vertices form an edge in $G$. Such a sequence is referred to as a \emph{path} between vertices $v_i$ and $v_j$. The shortest path is defined as 

\[ sp(v_i, v_j) = \argmin_p \left| p(v_i, v_j) \right| \]

and the length of the shortest path between $v_i$ and $v_j$ is known as the \emph{distance} between $v_i$ and $v_j$, denoted $d_{ij}$. The betweenness of a vertex $v_i$ is the number of shortest paths between any vertices that traverse vertex $v_i$:

\[ C_B(v_i) = \sum_{v_j, v_k \neq v_i} \frac{ \left| sp_i(v_j, v_k) \right|}{\left| sp(v_j, v_k) \right|} \]

where $sp_i(v_j, v_k)$ denotes a shortest path between vertices $v_j$ and $v_k$ traversing vertex $v_i$. 

\subsubsection{Closeness}
The closeness of the vertex $v_i$ is the average distance between the vertex $v_i$ and all other vertices in the network

\[ C_C(v_i) = \frac{1}{|V|} \sum\limits_{v_j \in V} d_{ij} \]

\subsubsection{Local clustering coefficient}
Local clustering coefficient of the vertex $v_i$ (also known as the \emph{transitivity} of $v_i$) measures the connectivity of the egocentric network of the vertex $v_i$. It is formally defined as the ratio of the number of edges existing in the egocentric network of the vertex $v_i$ to the maximum number of edges that could exist in this egocentric network. 

\[ C_L(v_i) = \frac{|E_i|}{|G_i|(|G_i-1|)} \]

\subsubsection{Eigencentrality}
The eigencentrality of the vertex $v_i$ is recursively defined as 

\[ C_E(v_i) = \frac{1}{\lambda} \sum\limits_{v_j \in V_i} C_E(v_j) \]

where $\lambda$ is a constant.

\subsection{Matrix Energies}

Depending on the properties of a particular matrix describing the network, several different energies with varying properties can be defined. In this research we are considering the following types of energies.

\subsubsection{Graph energy}

Graph energy is defined on the basis of the adjacency matrix $M_A$ of the network. Let

\[
    M_A[i,j] = 
    \begin{cases} 
        1 & \mathit{if} \ \{v_i,v_j\} \in E \\
        0 & \mathit{otherwise}
    \end{cases}
\]

be the \emph{adjacency matrix} of $G$. Then the graph energy of $G$ is defined as

\[ E_G(G) = \sum\limits_{i=1}^n |\mu_i| \]

where $\mu_1, \ldots, \mu_n$ are the eigenvalues of the adjacency matrix $M_A$. 

\subsubsection{Randi\'c energy}

Randi\'c matrix of the network is defined as:

\[ 
    M_R[i,j] = 
    \begin{cases} 
        0 & \mathit{if} \  i=j \\ 
        \frac{1}{\sqrt{C_D(v_i) C_D(v_j)}} & \mathit{if} \ \{v_i, v_j\} \in E \\ 
        0 & \mathit{if} \ \{v_i, v_j\} \notin E
    \end{cases} 
\]

Randi\'c energy is defined as 

\[ E_R(G) = \sum\limits_{i=1}^n |\rho_i| \] 

where $\rho_1, \ldots, \rho_n$ are the eigenvalues of the Randić matrix $M_R$.

\subsubsection{Laplacian energy}

The Laplacian matrix of a network is defined as:

\[ 
    M_L[i,j] = 
    \begin{cases} 
        d_i & \mathit{if} \ i=j \\ 
        -1 & \mathit{if} \ i \neq j \; \wedge \; \{v_i,v_j\} \in E \\ 
        0 & \mathit{otherwise}  
    \end{cases}
\]

Laplacian energy is defined as 

\[ E_L(G) = \sum\limits_{i=1}^n |\lambda_i-\frac{2m}{n}| \]

where $\lambda_1, \ldots, \lambda_n$ are the eigenvalues of the Laplacian matrix $M_L$, $m=|E|$ is the number of edges, and $n=|V|$ is the number of vertices. Laplacian energy of a network is equal to its Graph energy if and only if the network is regular. (i.e., when all vertices have equal degree). In addition, Laplacian energy obeys the following inequalities~\cite{Gutman2006}:

\begin{eqnarray}
    E_L(G)    & \leq & \sqrt{2Mn} \\
    E_L(G)    & \leq & \frac{2m}{n} + \sqrt{(n-1)\left[ 2M - (\frac{2m}{n})^2 \right]} \\
    2\sqrt{M} & \leq & E_L(G) \leq 2M 
    \label{eq:laplacian.energy.inequalities}
\end{eqnarray}

where $M = m + \frac{1}{2} \sum\limits_{i=1}^n (d_i - \frac{2m}{n})^2$ and $\frac{2m}{n}$ is the average vertex degree. 

\section{Network Models}
\label{sec:network.models}

\begin{figure}
    \centering
    \begin{subfigure}[b]{0.45\textwidth}
        \includegraphics[width=\textwidth]{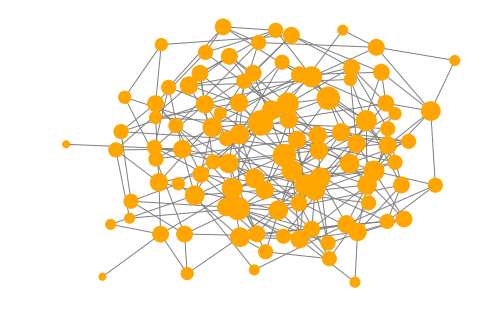}
        \caption{Erd\H{o}s-R\'enyi}
        \label{fig:erdos}
    \end{subfigure}
    ~ 
    \begin{subfigure}[b]{0.45\textwidth}
        \includegraphics[width=\textwidth]{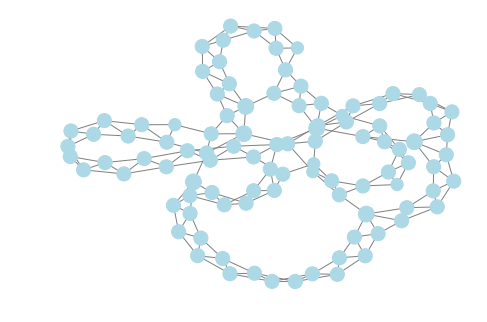}
        \caption{Watts-Strogatz}
        \label{fig:watts}
    \end{subfigure}
    ~
    \begin{subfigure}[b]{0.45\textwidth}
        \includegraphics[width=\textwidth]{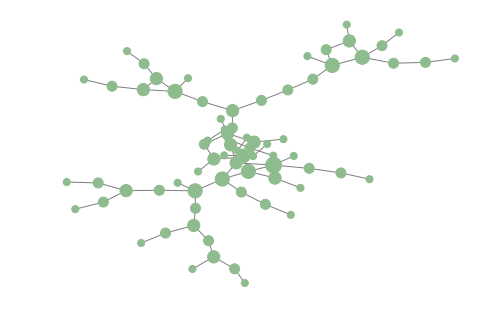}
        \caption{Waxman}
        \label{fig:waxman}
    \end{subfigure}
    ~ 
    \begin{subfigure}[b]{0.45\textwidth}
        \includegraphics[width=\textwidth]{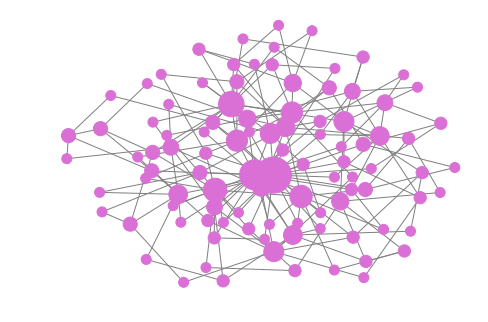}
        \caption{Holme-Kim}
        \label{fig:holme}
    \end{subfigure}
    \caption{Examples of networks produced by generative network models}\label{fig:networks}
\end{figure}

Over the years several generative network models have been proposed. Most of these models aim at generating artificial networks which display certain properties that are frequent in empirical networks. For instance, the prevalence of real-world social networks displays the power-law distribution of vertex degrees~\cite{clauset2009power}, and the Holme-Kim model aims at re-creating this property. Another frequent property of empirical networks is a very high local clustering coefficient~\cite{fagiolo2007clustering}, this property is prominent in the Watts-Strogatz model. In our experiments we have decided to use four different generative network models which cover a vast spectrum of possible network topologies, examples of networks produced by these models are presented in Figure~\ref{fig:networks}.

\subsection{Erd\H{o}s-R\'enyi random network model}

The random network model has been first introduced by Paul Erd\H{o}s and Alfr\'ed R\'enyi  in~\cite{erd6s1960evolution}. There are two versions of the model. The $G(n,m)$ model consists in randomly selecting a single network $g$ from the universe of all possible networks having $n$ vertices and $m$ edges. The second model, dubbed $G(n,p)$, creates the network $g$ by first creating $n$ isolated vertices, and then creating, for each pair of vertices $(v_i,v_j)$, an edge with the probability $p$. Due to much easier implementation and analytical accessibility, the $G(n,p)$ model is far more popular and this is the model which we have used in our experiments. 

\subsection{Watts-Strogatz small world network model}
The small world model has been introduced by Duncan Watts and Steven Strogatz in~\cite{Watts98}. According to this model, a set of $n$ vertices is organized into a regular circular lattice, with each vertex connecting directly to $k$ of its nearest neighbors. After creating the initial lattice, each edge is rewired with the probability $p$, i.e.\ an edge $(v_i,v_j)$ is replaced with the edge $(v_i,v_k)$ where $v_k$ is selected uniformly from $V$. 

\subsection{Holme-Kim preferential attachment network model}
There is a whole family of graph models collectively referred to as \emph{preferential attachment} or \emph{cumulative advantage} models. Basically, any model in which the probability of forming an edge to a vertex is proportional to that vertex degree can be classified as a preferential attachment model. The first attempt to use the mechanism of preferential attachment to generate artificial graphs can be attributed to Derek de Solla Price who tried to explain the process of formation of scientific citations by the advantageous accumulation of citations by prominent papers~\cite{de1976general}. Another well-known model which utilizes the same mechanism has been proposed by Albert-L\'aszl\'o Barab\'asi and R\'eka Albert~\cite{barabasi1999emergence}. According to this model, the initial graph is a full graph $K_{n_0}$ with $n_0$ vertices. All subsequent vertices are added to the graph one by one, and each new vertex creates $m$ edges. The probability of choosing the vertex $v$ as the endpoint of a newly created edge is proportional to current degree of $v$ and can be expressed as $p(v_i) = C_D(v_i) / \sum_j C_D(v_j)$. The resulting network displays the power law distribution of vertex degrees because of the quick accumulation of new edges by prominent vertices. In this work we are using the variation of the preferential attachment model proposed by Holme and Kim~\cite{holme2002growing}, in which after adding a random edge in the network, with a probability $p_\Delta$ an extra triangle closing edge is added to the network. This model is basically the same as the Barab\'asi-Albert model with an added triangle closure step which increases the local clustering coefficient.

\subsection{Waxman geometric random network model}
The Waxman geometric random network model~\cite{waxman1988routing} places $n$ vertices uniformly at random in a rectangular domain, two vertices $(v_i,v_j)$ are connected with an edge with probability $p_{ij} = \alpha*\exp(\frac{-d_{ij}}{\beta*d_{\mathit{max}}})$, where $d_{ij}$ is the Euclidean distance between vertices $v_i$ and $v_j$, and $d_{\mathit{max}}$ is the maximum distance between all vertices in the network (also known as the \emph{diameter} of the network). Parameters $\alpha$ and $\beta$ control the influence of the distance between any given pair of vertices in relation to the diameter of the network.

\section{Experiments}
\label{sec:experiments}

In the following section we present the results of the experimental evaluation of egocentric graph energies in different artificial networks. The experimental protocol is as follows. For each generative network model we generate several instances of the model, modifying a single model parameter. Then, we compute all the energies of all egocentric networks (of radius $r=1$) and we compute the entropy of the distribution of each energy. Thus, for each vertex, we add three features representing the Graph energy, the Randi\'c energy, and the Laplacian energy of that vertex egocentric network.

As we have noted before, generative network models try to mimic certain phenomena and mechanisms of network formation. It should be noted, however, that the topologies of networks produced by the same generative network model can vary dramatically depending on the choice of model's parameters. For instance, if we generate two networks of $n=100$ vertices each using the Erd\"os-R\'enyi random network model, but in the first network we set the random edge probability to $p=0.01$, and in the second network we set this parameter to $p=0.1$, the resulting networks will be very different in terms of their topology, distribution of degree, betweenness, eigencentrality, etc. In order to make sure that we control for the influence of the model's parameter, in our experiments we generate hundreds of instances of networks for each generative network model, each time slightly varying the value of the model's parameter.

The gradual changes of each model are realized by modifying the following parameters:

\begin{itemize}
    \item Erd\H{o}s-R\'enyi random network model: the probability of creating an edge between a random pair of vertices changes uniformly from $p=0.01$ to $p=1.0$, the network changes gradually from a set of isolated vertices to a fully connected network.
    \item Watts-Strogatz small world network model: the probability of randomly rewiring an edge changes uniformly from $p=0.01$ to $p=1.0$, the network changes gradually from a strictly ordered structure, where each vertex links to its $k=4$ nearest neighbors with no long-distance bridges, to a fully random network.
    \item Waxman geometric random network model: we change $\alpha$ uniformly in the range $[0.01, 1.0]$ while $\beta$ stays constant at $\beta=0.1$. 
    \item Holme-Kim powerlaw network model: the probability of closing a triangle after adding a random edge changes uniformly from $p=0.01$ to $p=1.0$, the resulting network is a scale-free network with power law degree distribution, subsequent instances of the network exhibit gradually increasing values of the average local clustering coefficient.
\end{itemize}

After generating each instance of a network, we compute, for each vertex, the following features and all correlations between these features:

\begin{itemize}
    \item degree
    \item betweenness
    \item closeness
    \item eigencentrality
    \item Graph energy of the ego-network of the vertex
    \item Randi\'c energy of the ego-network of the vertex
    \item Laplacian energy of the ego-network of the vertex
\end{itemize}

and we we show how these correlations change as the topology of each network model undergoes a gradual change. We use Pearson product-moment correlation coefficient to measure the similarity between features.

We begin by investigating the relationship between graph energies and vertex features in random networks.

\subsection{Erd\"os-R\'enyi random network model}

Below we present the correlations of graph energies with other vertex features. Horizontal x-axis represents the value of the model's parameter, namely, the random edge creation probability $p$. Please note that for low values of this probability, Erd\"os-R\'enyi random network model produces very sparse networks with many isolated components. 

\begin{figure}
    \centering
    \begin{subfigure}[b]{0.65\textwidth}
        \includegraphics[width=\textwidth]{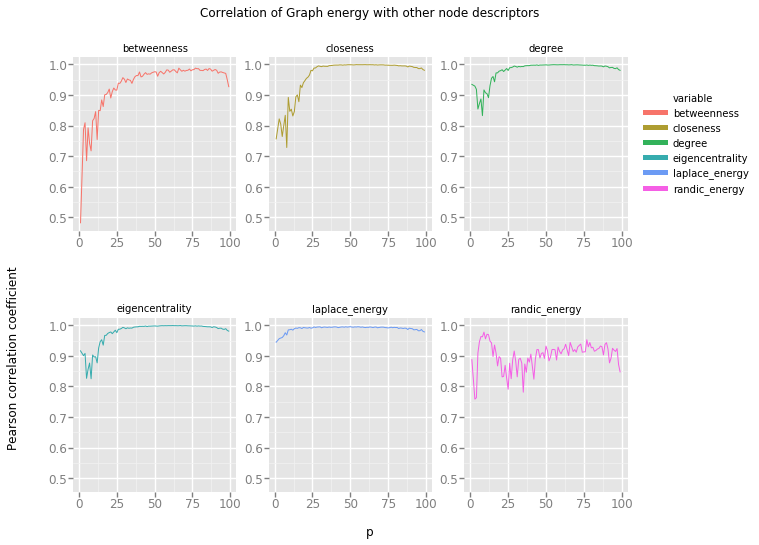}
        % \caption{Graph energy}
        \label{fig:random.graph.energy}
    \end{subfigure}
    ~ 
    \begin{subfigure}[b]{0.65\textwidth}
        \includegraphics[width=\textwidth]{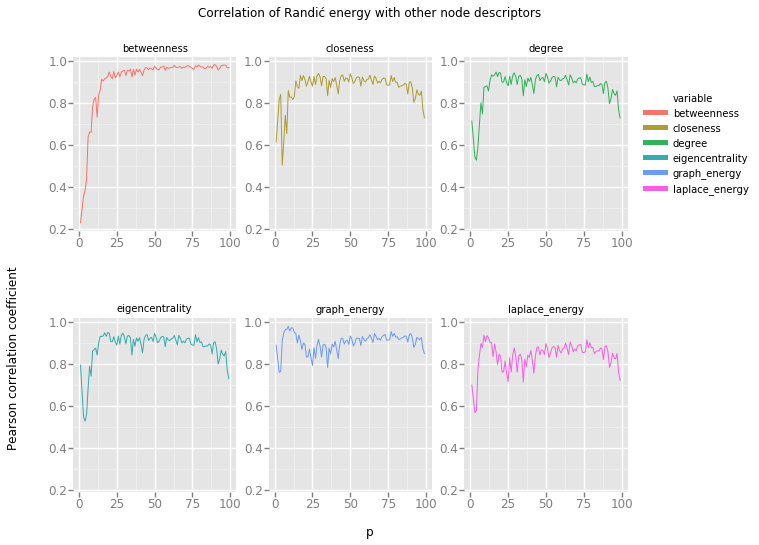}
        % \caption{Randi\'c energy}
        \label{fig:random.randic.energy}
    \end{subfigure}
    ~
    \begin{subfigure}[b]{0.65\textwidth}
        \includegraphics[width=\textwidth]{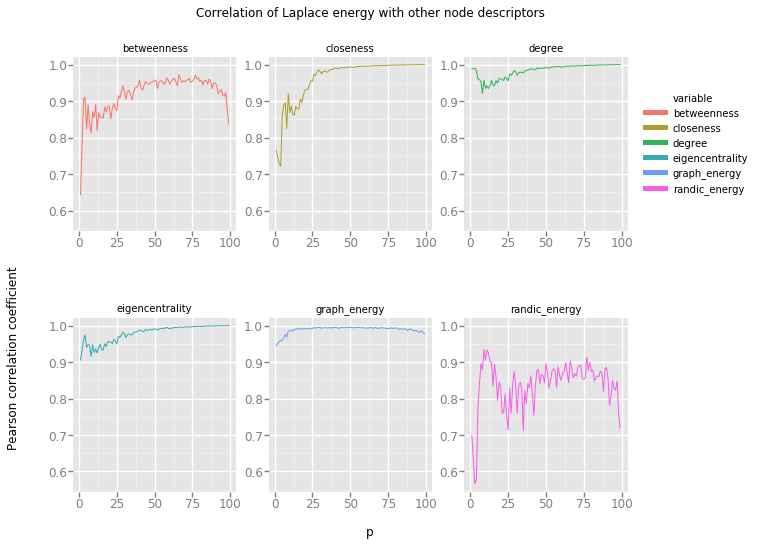}
        % \caption{Laplacian energy}
        \label{fig:random.laplacian.energy}
    \end{subfigure}
    \caption{Graph energies and the Erd\"os-R\'enyi random network model}
    \label{fig:erdos.energies}
\end{figure}

Our observations are as follows:

\begin{itemize}
    \item Graph energy correlates strongly with closeness and eigencentrality, even for relatively low probability of random edge creation ($p=0.2$), in these networks there is enough connectivity between vertices to produce large connected components, but the networks are far from being fully connected.
    \item Randić energy correlates well with betweenness, one can investigate the possibility of estimating the betweenness of a node based on its Randić energy for a wide spectrum of random networks.
    \item Laplacian energy correlates almost perfectly with degree, eigencentrality and closeness. Since both eigencentrality and closeness are expensive to compute, one can estimate these values based on the Laplacian energy of a vertex.
\end{itemize}

\subsection{Watts-Strogatz small world network model}

\begin{figure}
    \centering
    \begin{subfigure}[b]{0.65\textwidth}
        \includegraphics[width=\textwidth]{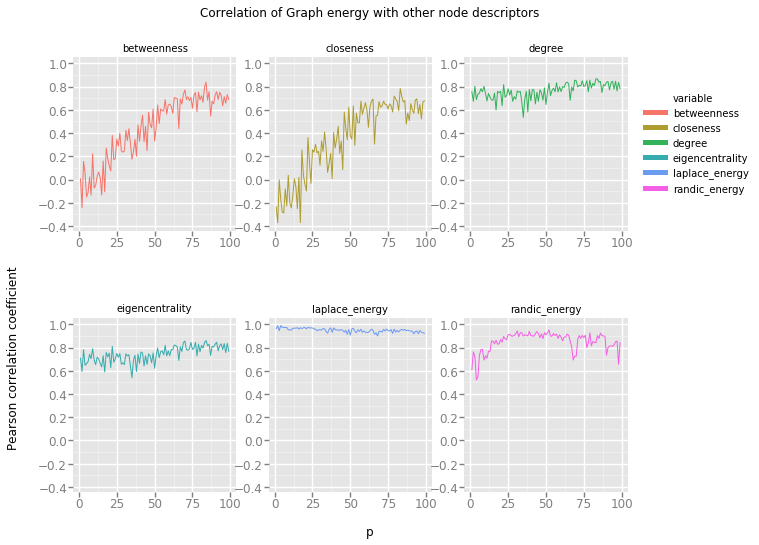}
        % \caption{Graph energy}
        \label{fig:smallworld.graph.energy}
    \end{subfigure}
    ~ 
    \begin{subfigure}[b]{0.65\textwidth}
        \includegraphics[width=\textwidth]{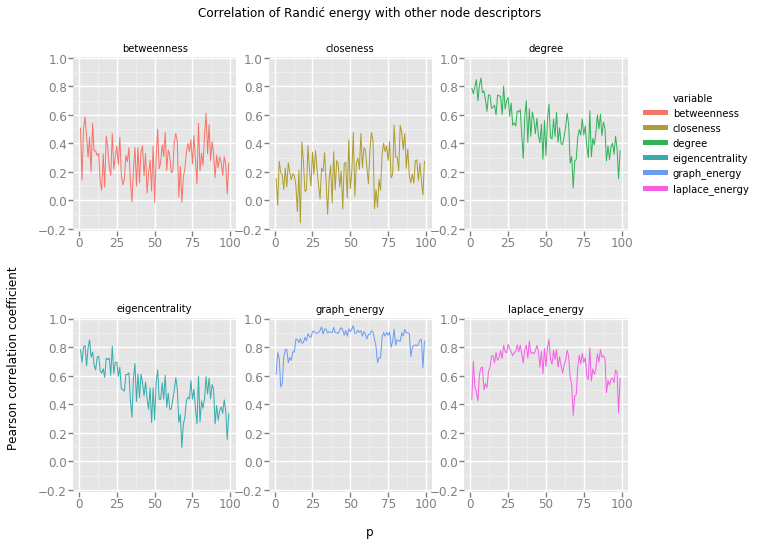}
        % \caption{Randi\'c energy}
        \label{fig:smallworld.randic.energy}
    \end{subfigure}
    ~
    \begin{subfigure}[b]{0.65\textwidth}
        \includegraphics[width=\textwidth]{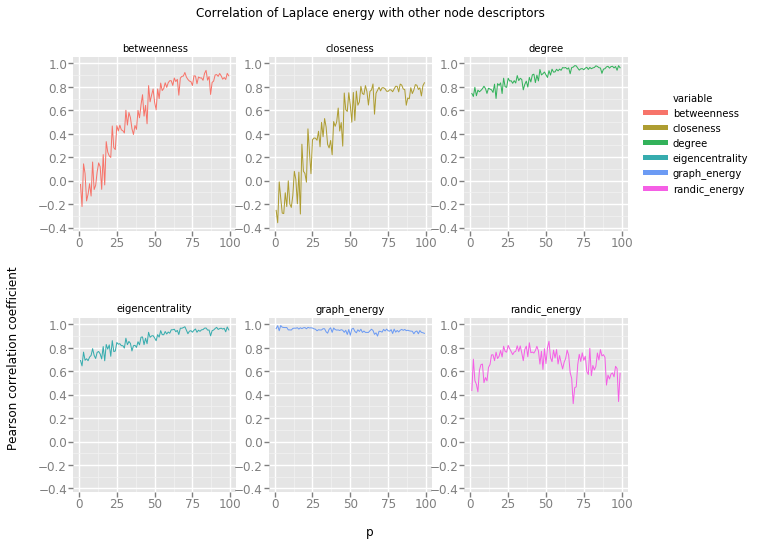}
        % \caption{Laplacian energy}
        \label{fig:smallworld.laplacian.energy}
    \end{subfigure}
    \caption{Graph energies and the Watts-Strogatz small world network model}
    \label{fig:watts.energies}
\end{figure}

Recall that in the Watts-Strogatz small world network model the main parameter, the probability $p$ of the random edge rewiring, determines the topology of the resulting network by gradually moving from a fully regular lattice of vertices to a fully random network. In Figure~\ref{fig:watts.energies} low values of the $p$ parameter on the x-axis represent very regular networks, and high values of the $p$ parameter correspond to irregular, random-like networks.

Looking at the results we note the following:

\begin{itemize}
    \item Graph energy correlates only with Laplacian energy, there is some weak correlation with degree and eigencentrality, but most probably this correlation is too weak to provide accurate estimates.
    \item Randić energy cannot be reasonably used to provide any estimates regarding the features of vertices in small world networks.
    \item Laplacian energy correlates to some extent with degree and eigencentrality, providing means for estimation, but this can be achieved only when the probability $p$ of random edge rewiring is sufficiently high.
\end{itemize}

\subsection{Waxman geometric random network model}

In general, Waxman geometric random network model is characterized by two parameters, $\alpha$ and $\beta$, which both determine the propensity of vertices to form short distance and long distance connections. In order to maintain consistency with the experimental protocol we have decided to set the constant value of $\beta=0.1$ and change the preference of vertices to create short distance connections. Low values of the $p$ parameter correspond to networks in which there is no explicit preference for short distance connections, and high values of the $p$ parameter produce networks in which the majority of connections are formed at short distances, making the resulting network more similar to the small world network model.

\begin{figure}
    \centering
    \begin{subfigure}[b]{0.65\textwidth}
        \includegraphics[width=\textwidth]{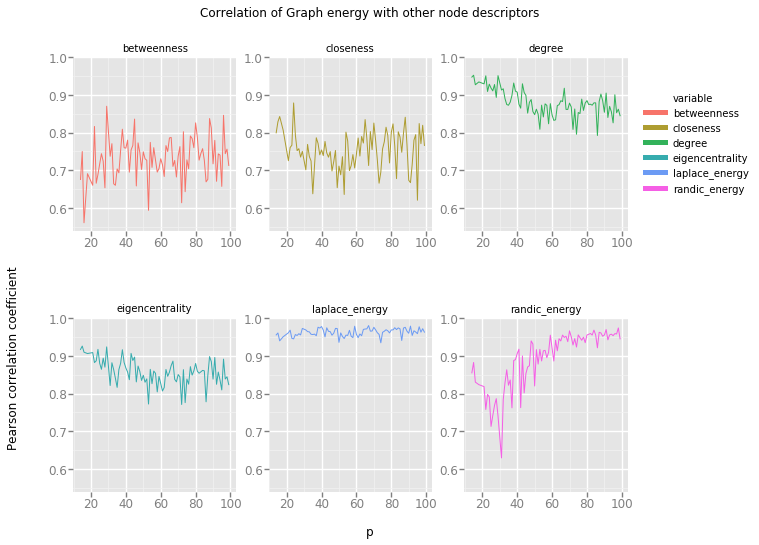}
        % \caption{Graph energy}
        \label{fig:waxman.graph.energy}
    \end{subfigure}
    ~ 
    \begin{subfigure}[b]{0.65\textwidth}
        \includegraphics[width=\textwidth]{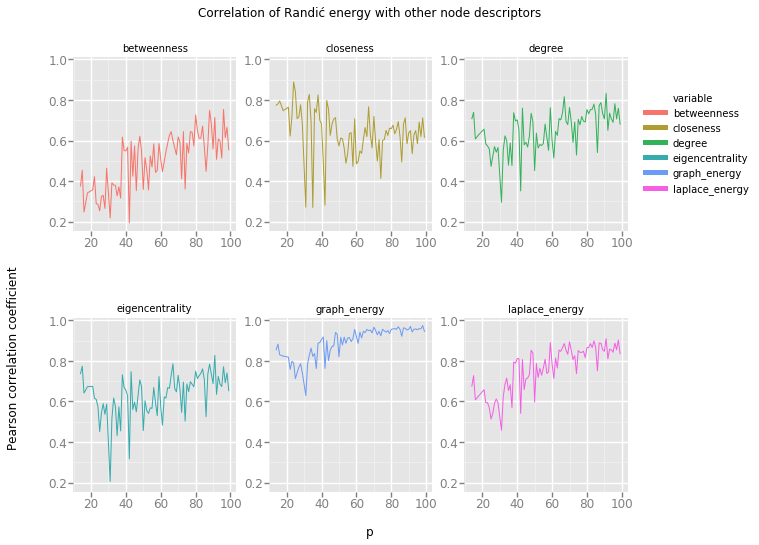}
        % \caption{Randi\'c energy}
        \label{fig:waxman.randic.energy}
    \end{subfigure}
    ~
    \begin{subfigure}[b]{0.65\textwidth}
        \includegraphics[width=\textwidth]{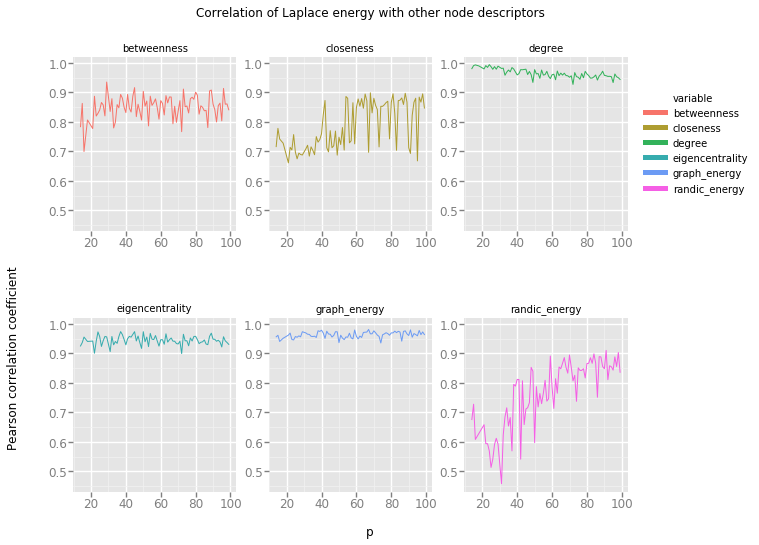}
        % \caption{Laplacian energy}
        \label{fig:waxman.laplacian.energy}
    \end{subfigure}
    \caption{Graph energies and the Waxman geometric random network model}
    \label{fig:waxman.energies}
\end{figure}

Our interpretation of the results is the following:

\begin{itemize}
    \item Graph energy correlates with degree and eigencentrality for small values of the $p$ parameter.
    \item Randić energy again does not seem to be very useful for estimating the values of other vertex features.
    \item Laplacian energy correlates very strongly with degree, and somehow strongly with eigencentrality. Most of these energies are quite stable across the spectrum of possible values of the $p$ parameter.
\end{itemize}

\subsection{Holme-Kim preferential attachment model}

\begin{figure}
    \centering
    \begin{subfigure}[b]{0.65\textwidth}
        \includegraphics[width=\textwidth]{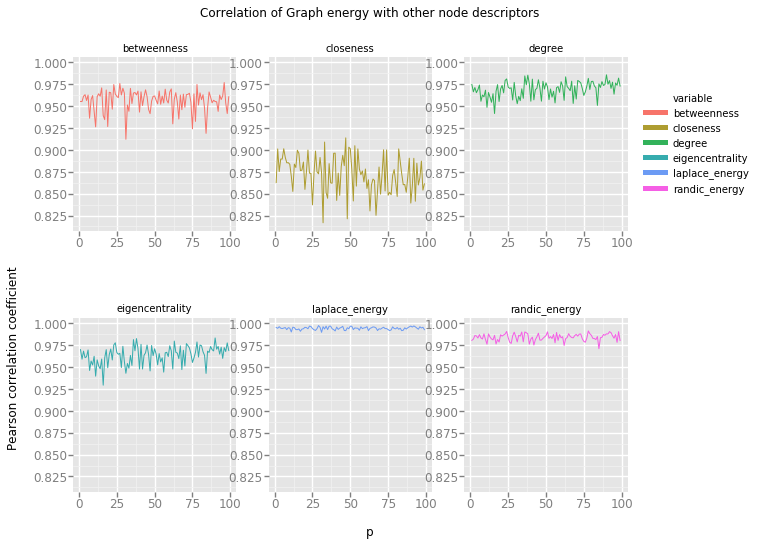}
        % \caption{Graph energy}
        \label{fig:powerlaw.graph.energy}
    \end{subfigure}
    ~ 
    \begin{subfigure}[b]{0.65\textwidth}
        \includegraphics[width=\textwidth]{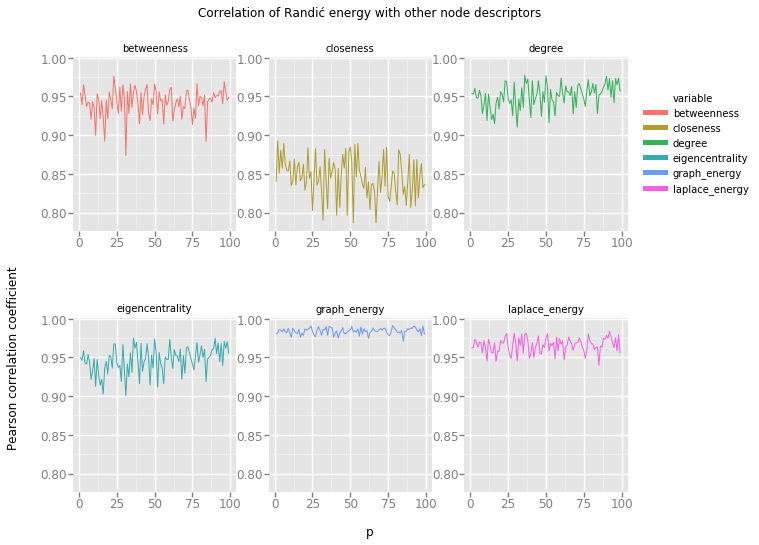}
        % \caption{Randi\'c energy}
        \label{fig:powerlaw.randic.energy}
    \end{subfigure}
    ~
    \begin{subfigure}[b]{0.65\textwidth}
        \includegraphics[width=\textwidth]{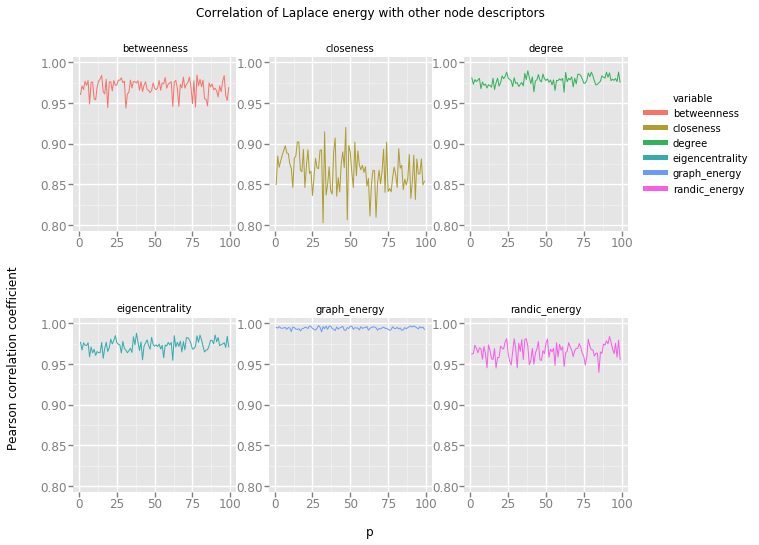}
        % \caption{Laplacian energy}
        \label{fig:powerlaw.laplacian.energy}
    \end{subfigure}
    \caption{Graph energies and the Holme-Kim preferential attachment  network model}
    \label{fig:powerlaw.energies}
\end{figure}

Figure~\ref{fig:powerlaw.energies} presents the correlation of graph energies with various vertex features in networks produced by the Holme-Kim preferential attachment network model. Recall that in this model the value of the parameter $p$ determines the propensity of vertices to close open triads. In other words, as the value of the parameter $p$ increases along the x-axis, the resulting networks become more and more clustered (making them more similar to small world networks), at the same time retaining the power law distribution of vertex degrees.

From the results we note, that:

\begin{itemize}
    \item Graph energy correlates with betweenness, degree and eigencentrality to the extent which allows us to suspect, that it is possible to estimate these descriptors based on the graph energy of the vertex.
    \item Randić energy correlates well with other types of vertex energies, but the correlations with degree, eigencentrality and betweenness are significant and can be useful.
    \item Laplacian energy exhibits very strong correlation with degree, betweenness and eigencentrality.
\end{itemize}

This is a very encouraging result. The Holme-Kim network model is a good approximation of many empirical networks, in particular, of large social networks. Very strong correlations of vertex features and graph energies are a clear indicator of the usefulness of the later in social network analysis.

\subsection{Stability of graph energies across possible network spectrum}

Another question worth answering is how stable are graph energies when different topologies of networks are considered. In our experiments we perform gradual change of each network, producing instances of slightly different topologies. In order to measure the degree to which energy disperses across the network, we use Shannon's information entropy. Low entropy of energy dispersion characterizes networks with uniform distribution of energy across vertices, and high entropy describes networks, in which certain areas of the network exhibit high variations of energy. 

% elaborate
This leads directly to the concept of energy gradients in networks, a research direction which we plan to undertake in the near future.

\begin{figure}
    \centering
    \includegraphics[width=0.95\textwidth]{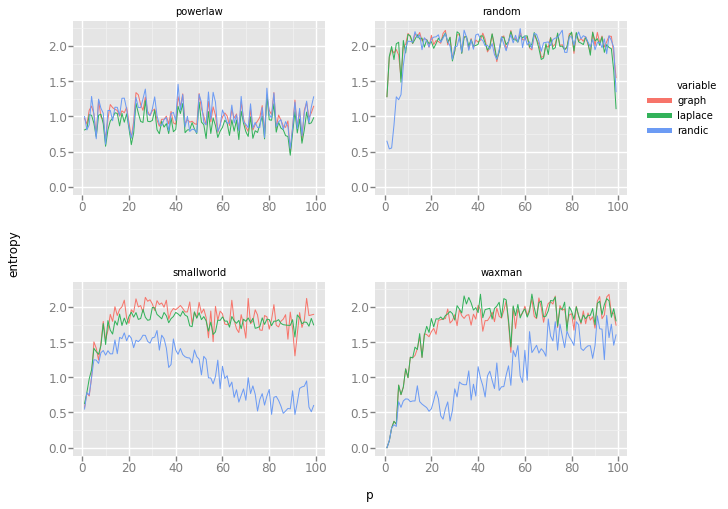}
    \caption{Entropies of graph energies for different generative network models}
    \label{fig:entropy}
\end{figure}

The results are presented in Figure~\ref{fig:entropy}. Not only does the entropy change for each generative network model, but there are visible differences between the models as well.

\begin{itemize}
    \item Erd\H{o}s-R\'enyi random network model: entropies of all energies quickly increase and stay at the maximum level during the densification of the graph, and only when the edge probability creation reaches 1 (leading to a single clique), the entropies drop to zero (as expected, because all vertices are now exactly the same and indistinguishable). Even in very dense random networks (for large values of the $p$ parameter) graph entropies are scattered across vertices with high variability.
    \item Watts-Strogatz small world network model: Interestingly, the addition of random rewired edges affects the entropies only at the beginning, but after reaching a certain threshold, the entropy begins to diminish. Initially, the entropies of graph energies are very low because the network is regular and all vertices are indistinguishable. Increase of energy entropies indicates the diversification of energy distribution among vertices. Interestingly, as the network becomes more "random" (i.e. more edges have been randomly rewired), the entropy of Randi\'c energy begins to steadily diminish (all vertices diverge to a common Randi\'c energy). This is probably caused by the fact, that initially all vertices form identical egocentric networks (with only minuscule variations), and as more and more edges becomes randomly rewired, these egocentric networks again become more unified. This is our conjecture which will require further examination.
    \item Waxman geometric random network model: graph energy and Laplace energy entropies behave similarly to the small world network model of Watts and Strogatz, but the entropy of Randi\'c energy steadily grows as the value of the $p$ parameter increases. This is really intriguing because, in theory, large values of the $p$ parameter should produce networks more similar to the small-world model, with the majority of edges formed between neighboring vertices. Increasing entropy suggests that Randi\'c energy becomes more dispersed among vertices. Without further investigation we cannot provide an informed explanation of this phenomenon.
    \item Holme-Kim powerlaw network model : entropies of all energies are constant across possible topologies of the model, with some random fluctuations. This is very much what we expect: changing the probability of triad closure (the increase of the $p$ parameter) does not change the shape of the distribution of vertex degrees in a significant way, this distribution is still best described using the powerlaw formula, irrespective of the triad closure probability.
\end{itemize}

\section{Conclusions}
\label{sec:conclusions}

In this paper we have presented the first investigation of the properties of various graph energies in the field of complex networks. Graph energy is the energy of a symmetric matrix representing a network. Since there are several different types of matrices which can be used to describe a network (e.g., an adjacency matrix, a distance matrix, a Laplacian), each of these matrices produces different energy. When applied to the matrix representation of an entire network, graph energy has limited information value and equivocal interpretation. One could argue that networks with high energies have large overall capacity, i.e., they allow for stable allocation of large amounts of resources due to the existence of a stationary distribution of that allocation. This hypothesis requires further scientific investigation. It should be clear that the usefulness of graph energy in describing the topology of chemical compounds or other relatively small graphs does not transfer to the realm of large complex networks easily. However, we discover that graph energies applied to egocentric networks of individual vertices produce very meaningful and interesting results. Graph energy of an egocentric network for a vertex is an explainable and interpretable measure of the capacity of that vertex. Graph energy, in our opinion, provides a characterization of a vertex which is complementary to topological characteristics (degree, eigencentrality) and distance-based characteristics (betweenness, closeness, eccentricity). In particular, we observe strong correlation of graph energy and other descriptors such as betweenness and eigencentrality for some generative network models. This is very exciting, because these vertex features are notoriously difficult to compute as it is impossible to compute them locally. For instance, computing vertex betweenness involves finding all shortest paths between all pairs of vertices, an operation which can be computationally prohibitively expensive for large complex networks. Our findings allow us to pursue promising direction involving  estimation of betweenness and eigencentrality from local vertex graph energies.

\acknowledgments{This work was supported by the National Science Centre, Poland, decision no. DEC-2016/23/B/ST6/03962.}

\authorcontributions{Both authors conceived and designed the experiments, performed the experiments, analyzed the data, and  wrote the manuscript equally.}

\conflictofinterests{The authors declare no conflict of interest.}

\end{document}